\documentclass[aps, prl, twocolumn,amsmath,amssymb, longbibliography]{revtex4-1}
\usepackage{graphicx}
\usepackage{dcolumn}
\usepackage{bm}
\usepackage{textcomp,mathcomp}
\usepackage{fancyhdr}
\usepackage{natbib}

\begin{document}

\title{Macroscopic quantum oscillator based on a flux qubit}

\author{Mandip Singh}
\email{mandip@iisermohali.ac.in}
\affiliation{Department of Physical Sciences
\\Indian Institute of Science Education and Research Mohali, Sector-81, S.A.S. Nagar, Manauli 140306,
India.}

\begin{abstract}
In this paper a macroscopic quantum oscillator is proposed, which consists of a flux-qubit in the form of a cantilever. The net magnetic flux threading through the flux-qubit and the mechanical degrees of freedom of the cantilever are naturally coupled. The coupling between the cantilever and the magnetic flux is controlled through an external magnetic field. The ground state of the flux-qubit-cantilever turns out to be an entangled quantum state, where the cantilever deflection and the magnetic flux are the entangled degrees of freedom. A variant, which is a special case of the flux-qubit-cantilever without a Josephson junction, is also discussed.
\end{abstract}
\maketitle

\section{Introduction}
   The experimental realization of quantum superposition and quantum entanglement of macroscopically distinct quantum states or Schr\"{o}dinger cat states is one of the main objectives of experiments to explore quantum physics at a macroscopic scale \cite{legget, ques}. The magnetic flux quantization signifies a quantum effect at a mesoscopic scale, where a net magnetic flux threading through a closed superconducting loop is quantized. From a fundamental point of view, a quantum superposition of distinct magnetic flux states is discussed in reference \cite{legget}. Experimentally, a quantum superposition of distinct magnetic flux states has been realized, based on a superconducting loop interrupted by Josephson junctions (flux-qubit) \cite{fried,caspar, poletto}. Flux-qubits, sometimes referred to as ``designed macroscopic atoms", have been studied theoretically and experimentally in the context of quantum information processing \cite{devoret, revwendin, mooij}. On the other hand, a nano or a micromechanical cantilever \cite{Mo}, which is regarded as a macroscopic quantum harmonic oscillator, can exhibit quantum behaviour at low temperatures. Such a cantilever, when strongly coupled to another quantum system, can exhibit quantum entanglement at the macroscopic scale. Various approaches are being explored through experiments to strongly couple a nano or a micromechanical cantilever to another quantum system, such as photons in a cavity \cite{penrose, mirrorcooling, selfcooling, membrane, photonics, qresonator, qresonator2, rob}, nitrogen vacancies \cite{nitrogen},   a Bose-Einstein condensate \cite{bec} and superconducting quantum circuits \cite{Lahaye, phonon, mesoscale, sqgnd, nori, you, neill, fran, hybr}.

In this paper, a macroscopic quantum oscillator based on a flux-qubit is proposed. The quantum oscillator, which is named as the flux-qubit-cantilever, consists of a flux-qubit where a part of the flux-qubit loop is in the form of a cantilever. The magnetic flux threading through the flux-qubit and the mechanical degree of freedom of the cantilever are naturally coupled to each other, and the coupling is controlled through an external magnetic field. The potential energy profile of the flux-qubit-cantilever is tunable from a two-dimensional single well potential to a two-dimensional double well potential by appropriately tuning the magnetic field and the cantilever equilibrium angle. The resulting ground state of the flux-qubit-cantilever is an entangled quantum state of the magnetic flux and the deflection of the cantilever. In a proposal by Xue $et.al$ \cite{xue1, xue2} the coupling of a nanobeam mechanical resonator embedded in a superconducting circuit is controlled with an external magnetic field. An experimental realization of such a mechanical resonator embedded in a dc-SQUID is described in reference \cite{etaki} where the thermal motion of a micromechanical resonator is experimentally measured at milliKelvin temperatures. Semiclassical dynamics of a nanomechanical oscillator coupled to a rf-SQUID is described in reference \cite{ella}. The current proposal is based on a generic approach and it has been proved in this paper that the ground state of the flux-qubit-cantilever is an entangled quantum state of macroscopic observables. In addition, a special case of the flux-qubit-cantilever without Josephson junctions, which is called a superconducting-loop-oscillator, is also discussed. The potential energy of this superconducting-loop-oscillator can be controlled by an external magnetic field to produce a macroscopic quantum superposition. Therefore, the proposal presented in this paper is distinct from and goes beyond the proposals described in references \cite{xue1, xue2}. More specifically, the following aspects are novel in the present proposal: (a) The coupling between the flux-qubit and the cantilever is obtained from general energy considerations. (b) All of the relevant variables are treated quantum mechanically. (c) It is clearly established that for desirable parameters, the ground state of the flux-qubit-cantilever is an entangled state of macroscopic observables. (d) For the case of superconducting-loop-oscillator (a special case of the flux-qubit-cantilever), the ground state is a quantum superposition of macroscopically distinct configurations.

The phenomenon of flux quantization is a consequence of the Aharanov-Bohm effect.
The persistent current flowing through a superconducting loop is proportional to $\nabla\varphi(r)-(2e/\hbar) A(r)$, where $\varphi(r)$ is the phase of the Cooper pair wavefunction, $A(r)$ is the vector potential, $e$ is the electron charge and $h$ is Plank's constant ($\hbar= h/2\pi$). Due to the Meissner effect, the persistent current decreases exponentially as one moves inwards from the surface of a superconductor. Therefore, the phase gradient is $\nabla\varphi(r)=(2e/\hbar)A(r)$ at a point $r$ situated within the superconductor at a distance much greater than the penetration depth from the surface. To maintain a single valued Cooper pair wavefunction, the total phase accumulated over a closed path (situated within the loop and enclosing the centre of the loop), has to be an integral multiple of $2\pi$ which results in the quantization of the net magnetic flux threading through the superconducting loop \cite{leggetrev}.

For a superconducting loop interrupted by a Josephson junction, a phase $\Delta\varphi$ is accumulated across the Josephson junction. Therefore, the net phase accumulated around a closed path (located within the superconducting loop and passing through the Josephson junction while enclosing the loop centre), is $\Delta\varphi+2\pi\Phi/\Phi_{o}$, where $\Phi$ is the net magnetic flux threading through the superconducting loop and $\Phi_{o}=h/2e$ is the magnetic flux quantum. To maintain continuity of the Cooper pair wavefunction, the phase accumulated around a closed path  should be an integral multiple of $2\pi$ such that $\Delta\varphi+2\pi\Phi/\Phi_{o}=2k\pi$ (where $k$ is an integer). The potential energy of a superconducting loop interrupted by a single Josephson junction (flux-qubit) consists of two components. The first component is the magnetic energy stored in the superconducting loop due to the magnetic flux $\Phi$ threading through it in the presence of an external magnetic flux $\Phi_{a}$ and the second component is the potential energy accumulated by the Cooper pairs while tunneling through the Josephson junction. The Josephson junction also forms a junction capacitor, however, for the flux-qubit the energy contribution due to energy stored at the junction capacitor is considered to be much less than the energy. Therefore, the effective Hamiltonian of the flux-qubit is \cite{leggetrev,revwendin}
\begin{equation}
\label{eq:potential}
H_{Q}=\frac{p^{2}_{\Phi}}{2 C}+\frac{(\Phi-\Phi_{a})^{2}}{2 L}+E_{j}\left(1-\cos(2\pi\Phi/\Phi_{o})\right)
\end{equation}
where $p_{_{\Phi}}=-i\hbar\partial/\partial\Phi$ is the momentum conjugate to $\Phi$, and the second term $(\Phi-\Phi_{a})^{2}/2 L$ is the magnetic energy stored in the flux-qubit loop of self-inductance $L$ for an external applied flux $\Phi_{a}$. The third term $E_{j}(1-\cos(2\pi\Phi/\Phi_{o})$ is the potential energy of the Josephson junction with Josephson energy $E_{j}=I_{c}\hbar/2e$ and  $I_{c}$ being the critical current \emph{i.e.} the maximum current that can pass through the Josephson junction without dissipation. The potential energy of the flux-qubit (near the flux bias point $\Phi_{a}$) corresponds to a symmetric one-dimensional double well if the qubit is biased at half of the flux quantum \emph{i.e} $\Phi_{a}=\Phi_{0}/2$.

\section{Flux-Qubit-Cantilever}
\begin{center}
\begin{figure}
\begin{center}
\includegraphics[scale=0.34]{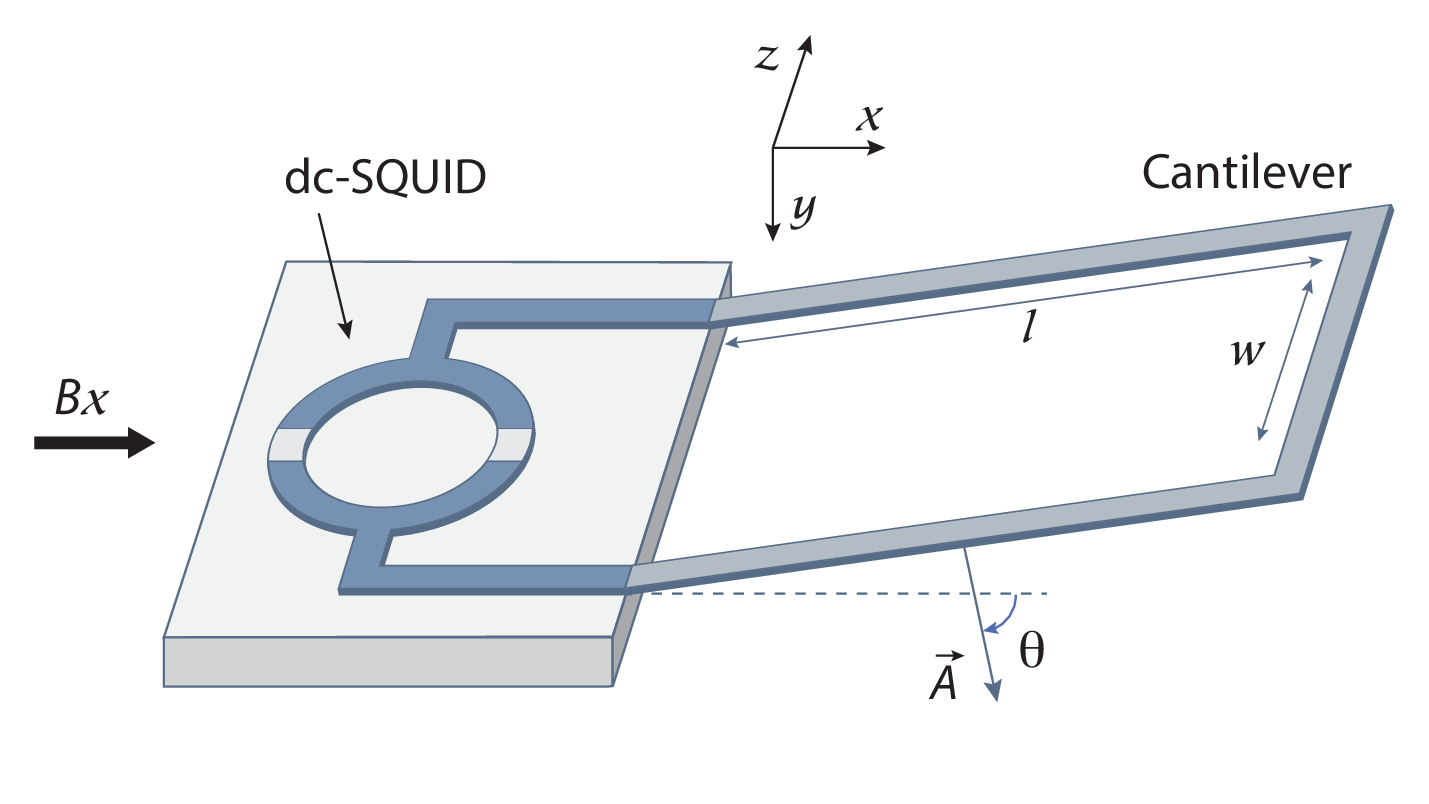}
\caption{\label{fig1} \emph{A schematic of the flux-qubit-cantilever. A part of the flux-qubit (larger loop) is projected from the substrate to form a cantilever. The external magnetic field $B_{x}$ controls the coupling between the flux-qubit and the cantilever. An additional magnetic flux threading through a dc-SQUID (smaller loop) which consists of two Josephson junctions adjusts the tunneling amplitude. The dc-SQUID can be shielded from the effect of $B_{x}$.}}
\end{center}
\end{figure}
\end{center}

Consider a part of the flux-qubit loop which is made to project from the substrate and the projected part acts as a cantilever and thereby provides us with an additional degree of freedom in the system. A schematic diagram of the flux-qubit-cantilever, where a part of the superconducting loop of a flux-qubit forms a cantilever, is shown in Fig.~\ref{fig1}.  In this diagram the larger loop is interrupted by a smaller loop consisting of a dc-SQUID. The Josephson energy that is constant for a single Josephson junction can be varied by applying a magnetic flux to the dc-SQUID loop. However, for the calculations throughout in this paper a flux-qubit with a single Josephson junction is considered. The external magnetic flux applied to the cantilever is $\Phi_{a}= B_{x} A \cos(\theta)$, where $B_{x}$ is the magnitude of the uniform external magnetic field along a fixed $x$-axis and the angle $\theta$ is the angle between the magnetic field ($x$-axis) and the area vector $\vec{A}$. Consider that the cantilever oscillates about an equilibrium angle $\theta_{0}$ with an intrinsic frequency of oscillation $\omega_{i}$ (the frequency in absence of the magnetic field). Since the external magnetic flux applied to the flux-qubit-cantilever depends on the cantilever deflection, therefore, the flux-qubit whose potential energy depends on an external flux is coupled to the cantilever. The potential energy of the flux-qubit-cantilever corresponds to a two-dimensional function $V(\Phi,\theta)$. Therefore, the Hamiltonian of the flux-qubit-cantilever interrupted by a single Josephson junction can be written as
\begin{eqnarray}\nonumber
\label{eq:hamiltonian}
H&=&\frac{p^{2}_{\theta}}{2 I_{m}}+\frac{1}{2}I_{m}\omega^{2}_{i}(\theta-\theta_{0})^{2} +\frac{p^{2}_{\Phi}}{2 C} \\&& +\frac{(\Phi-B_{x}A\cos(\theta))^{2}}{2 L}+E_{j}\left(1-\cos(2\pi\Phi/\Phi_{o})\right)
\end{eqnarray}
 The first two terms of Eq.~\ref{eq:hamiltonian} correspond to the Hamiltonian of the cantilever, the last three terms correspond to the Hamiltonian of the flux-qubit and its coupling to the cantilever. The fourth term is the magnetic energy of a superconducting loop in presence of an external magnetic field and the fifth term is the potential energy of the Josephson junction. Here $\Phi$ is the magnetic flux threading through the superconducting loop of the flux-qubit, $p_{\theta}=-i\hbar\partial/\partial\theta$ is the momentum conjugate to $\theta$, $I_{m}$ is the moment of inertia of the cantilever about the $z$-axis, $A$ is the area of the cantilever ($A=l\times w$ with $l$ being the length and $w$ being the width of the cantilever as shown in Fig.~\ref{fig1}). In case of $B_{x}=0$, the coupling between the mechanical degree of freedom of the cantilever and the magnetic flux of the flux-qubit is zero. For $\omega_{i}=0$, the cantilever has zero intrinsic restoring torque about the $z$-axis; however due to the magnetic energy stored in the loop and potential energy of the Josephson junction there exist several potential energy minima forming a two-dimensional multi-well potential.
 The two-dimensional potential energy of the flux-qubit-cantilever,
 $V(\Phi,\theta)=
 (\Phi-B_{x}A\cos(\theta))^{2}/2L + E_{j}\left( 1-\cos(2\pi\Phi/\Phi_{o}) \right)+I_{m}\omega^{2}_{i}(\theta-\theta_{0})^{2}/2$ has a single global minimum situated at ($n\Phi_{0},\theta^{+}_{n}$) if the equilibrium position of the cantilever is chosen to be $\theta_{0}=\theta^{+}_{n}$, where $\theta^{+}_{n}=+\cos^{-1}[n\Phi_{o}/B_{x}A]$, $m\Phi_{o} < B_{x}A < (m+1)\Phi_{o}$, integer $m \geq 0$ and  $n=-m,-m+1,..0..,m-1,m$. Here the integer $n$ is considered to be positive if the direction of the net magnetic field passing through the cantilever loop points along the area vector of the loop. The potential energy near the global minimum point is considered to be a single two-dimensional well. Similarly, a global minimum of $V(\Phi,\theta)$ is situated at ($n\Phi_{0},\theta^{-}_{n}$) if the equilibrium position of the cantilever is chosen to be $\theta_{0}=\theta^{-}_{n}$, where $\theta^{-}_{n}=-\cos^{-1}[n\Phi_{o}/B_{x}A]$. The location of the potential energy minimum depends on the cantilever equilibrium angle and it can be further fine-tuned by tilting the magnetic field direction \emph{w.r.t} the $x$-axis. Therefore, from the Taylor expansion of $V(\Phi,\theta)$ (up to second order differentiation) around a potential minimum $(n\Phi_{o},\theta^{+}_{n})$ of a single two-dimensional well the Hamiltonian given in Eq.~\ref{eq:hamiltonian} is written as
 \begin{eqnarray} \nonumber
\label{eq:taylor}
H_{n}&\simeq &\frac{p^{2}_{\phi}}{2 C}+\frac{p^{2}_{\delta}}{2 I_{m}}+\left(\frac{1}{2L} +\frac{2\pi^{2} E_{j}}{\Phi^{2}_{o}}\right)\phi^{2} \\&& \nonumber +\left(\frac{B^{2}_{x} A^{2}-n^{2}\Phi^{2}_{o}}{2 L}+\frac{1}{2}I_{m}\omega^{2}_{i}\right)~\delta^{2}\\&& + \frac{(B^{2}_{x} A^{2}-n^{2}\Phi^{2}_{o})^{1/2}}{L}~\phi \delta
 \end{eqnarray}
 Where the angle $\delta=\theta-\theta^{+}_{n}$ and the magnetic flux $\phi=\Phi-n\Phi_{o}$ are defined \emph{w.r.t} the two-dimensional potential well minimum $(n\Phi_{o},\theta^{+}_{n})$. The momentum conjugate to $\phi$ and $\delta$ are $p_{\phi}=-i\hbar\partial/\partial \phi$ and $p_{\delta}=-i\hbar\partial/\partial \delta$, respectively. The Hamiltonian in Eq.~\ref{eq:taylor} corresponds to a Hamiltonian of two coupled quantum harmonic oscillators of non-identical masses and different spring constants. The last term, containing a product of $\phi$ and $\delta$, represents the coupling between the oscillators. Due to this coupling term the eigen states of the Hamiltonian in Eq.~\ref{eq:taylor} cannot be written as a product of the magnetic flux states and the cantilever oscillator states. In other words, the eigen functions of the Hamiltonian given in Eq.~\ref{eq:taylor} are non-separable functions of $\phi$ and $\delta$. On the other hand, if the cantilever equilibrium angle is chosen to be $\theta_{0}=\theta^{-}_{n}$ then the global potential energy minimum is located at ($n\Phi_{o},\theta^{-}_{n}$) and the sign of the coupling term in Eq.~\ref{eq:taylor} is reversed.

 The Hamiltonian in Eq.~\ref{eq:taylor} is rewritten as
 \begin{eqnarray}
\label{eq:taylor2}
H_{n}&\simeq&\frac{p^{2}_{\phi}}{2 C}+\frac{p^{2}_{\delta}}{2 I_{m}}+\frac{1}{2}C\omega^{2}_{\phi}\phi^{2}+\frac{1}{2}I_{m}\omega^{2}_{\delta}\delta^{2}+\kappa\phi\delta
 \end{eqnarray}
 where oscillation frequencies along $\phi$ and $\delta$ are
 \begin{align}
 \label{eq:omega}
\omega^{2}_{\phi}&= \frac{1}{C}\left(\frac{1}{L}+\frac{4\pi^{2}E_{j}}{\Phi^{2}_{o}}\right)& \omega^{2}_{\delta}&= \left(\frac{B^{2}_{x}A^{2}-n^{2}\Phi^{2}_{o}}{I_{m}L}+\omega^{2}_{i}\right)
\end{align}
 and the coupling constant $\kappa$ (for $B_{x}A>n\Phi_{o}$) is
  \begin{align}
  \label{eq:kappa}
\kappa=\frac{(B^{2}_{x}A^{2}-n^{2}\Phi^{2}_{o})^{1/2}}{L}
\end{align}
The coupling constant $\kappa$ increases with the external magnetic field $B_{x}$. It is important to note that, even if the intrinsic frequency $\omega_{i}$ of the cantilever is zero, the cantilever experiences a restoring force in the presence of an external magnetic field that results in a nonzero $\omega_{\delta}$.

Consider a transformation of variables defined through the equations given below
\begin{align}
X&= \left(\frac{C}{I_{m}}\right)^{1/4}\cos(\beta) \phi+\left(\frac{I_{m}}{C}\right)^{1/4} \sin(\beta) \delta &
\\ Y&=-\left(\frac{C}{I_{m}}\right)^{1/4} \sin(\beta) \phi +\left(\frac{I_{m}}{C}\right)^{1/4} \cos(\beta) \delta
\end{align}
In terms of the new variables $X$ and $Y$, the Hamiltonian given in Eq.~\ref{eq:taylor2} transforms to a Hamiltonian of two uncoupled oscillators of identical masses $\mu=(C I_{m})^{1/2}$, such that
\begin{equation}
\label{eq:uncoupled}
H_{n}=\frac{P^{2}_{X}}{2\mu}+\frac{P^{2}_{Y}}{2\mu}+\frac{1}{2}\mu\omega^{2}_{X} X^{2}+\frac{1}{2}\mu\omega^{2}_{Y}Y^{2}
\end{equation}
for an angle of rotation
\begin{equation}
\label{eq:angle}
\beta=\frac{1}{2}\tan^{-1}\left[\frac{2\kappa/\mu}{\omega^{2}_{\phi}-\omega^{2}_{\delta}}\right]
\end{equation}
where $P_{X}=-i\hbar\partial/\partial X$ and $P_{Y}=-i\hbar\partial/\partial Y$ are the momenta conjugate to $X$ and $Y$, respectively. The eigen frequencies of the uncoupled Hamiltonian in Eq.~\ref{eq:uncoupled} are
\begin{equation}
\label{eq:eigenfreqx}
\omega^{2}_{X}=\omega^{2}_{\phi} \cos^{2}(\beta)+\omega^{2}_{\delta}\sin^{2}(\beta)+\frac{\kappa}{\mu} \sin(2\beta)
\end{equation}
and
\begin{equation}
\label{eq:eigenfrey}
\omega^{2}_{Y}=\omega^{2}_{\phi} \sin^{2}(\beta)+\omega^{2}_{\delta}\cos^{2}(\beta)-\frac{\kappa}{\mu} \sin(2\beta)
\end{equation}
The ground state wavefunction of the uncoupled Hamiltonian given in Eq.~\ref{eq:uncoupled} turns out to be
\begin{eqnarray}
\label{eq:groundxy}
\Psi^{n}_{0}(X,Y)&=&\left(\frac{\mu^{2}\omega_{X}\omega_{Y}}{\pi^{2}\hbar^{2}}\right)^{1/4} \\&& \nonumber \times \mathrm{exp}\left(-\frac{\mu \omega_{X}X^{2}}{2\hbar}\right)\mathrm{exp}\left(-\frac{\mu \omega_{Y}Y^{2}}{2\hbar}\right)
\end{eqnarray}
The above ground state wavefunction re-expressed in terms of variables $\phi$-$\delta$ is
\begin{eqnarray}
\label{eq:groundpd}
\Psi^{n}_{0}(\phi,\delta)&=&\left(\frac{C I_{m} \omega_{X}\omega_{Y}}{\pi^{2}\hbar^{2}}\right)^{1/4} \\&& \nonumber \times \mathrm{exp}\left(-\frac{C}{2\hbar}(\omega_{X}\cos^{2}(\beta)+\omega_{Y}\sin^{2}(\beta))~\phi^{2}\right) \\&& \times \nonumber \mathrm{exp}\left(-\frac{I_{m}}{2\hbar}(\omega_{X}\sin^{2}(\beta)+\omega_{Y}\cos^{2}(\beta))~\delta^{2}\right) \\&& \nonumber \times
\mathrm{exp}\left(-\frac{(C I_{_{m}})^{1/2}}{2\hbar}(\omega_{X}-\omega_{Y})\sin(2\beta)~\phi~\delta\right)
\end{eqnarray}

The last exponent of Eq.~\ref{eq:groundpd} consists of a product of $\phi$ and $\delta$, therefore, the ground state wavefunction $\Psi^{n}_{0}(\phi,\delta)$ is non-separable \emph{i.e.} $\Psi^{n}_{0}(\phi,\delta)$ cannot be written as a product of functions $\psi^{n}_{0}(\phi)$ and $\psi^{n}_{0}(\delta)$, where $\psi^{n}_{0}(\phi)$ and $\psi^{n}_{0}(\delta)$ are functions of $\phi$ and $\delta$ alone, respectively. The ground state wavefunction $\Psi^{n}_{0}(\phi,\delta)$ is separable if the coupling constant $\kappa = 0$.

The ground state $|\alpha\rangle^{n}_{0}$ of the flux-qubit-cantilever in the basis $|\phi\rangle|\delta\rangle$ is
\begin{eqnarray}
\label{eq:grounda}
|\alpha\rangle^{n}_{0}=\int\int C^{n}_{\phi,\delta} |\phi\rangle|\delta\rangle\mathrm{d}\phi\mathrm{d}\delta
\end{eqnarray}
where $C^{n}_{\phi,\delta}=(\langle\delta|\langle\phi|)|\alpha\rangle^{n}_{0}=\Psi^{n}_{0}(\phi,\delta)$. Since the ground state wave-function $\Psi^{n}_{0}(\phi,\delta)$ of the flux-qubit-cantilever is non-separable, therefore, the ground state $|\alpha\rangle^{n}_{0}$ is an entangled state.
\subsection{Experimental Considerations}
\begin{center}
\begin{figure}
\begin{center}
\includegraphics[scale=0.24]{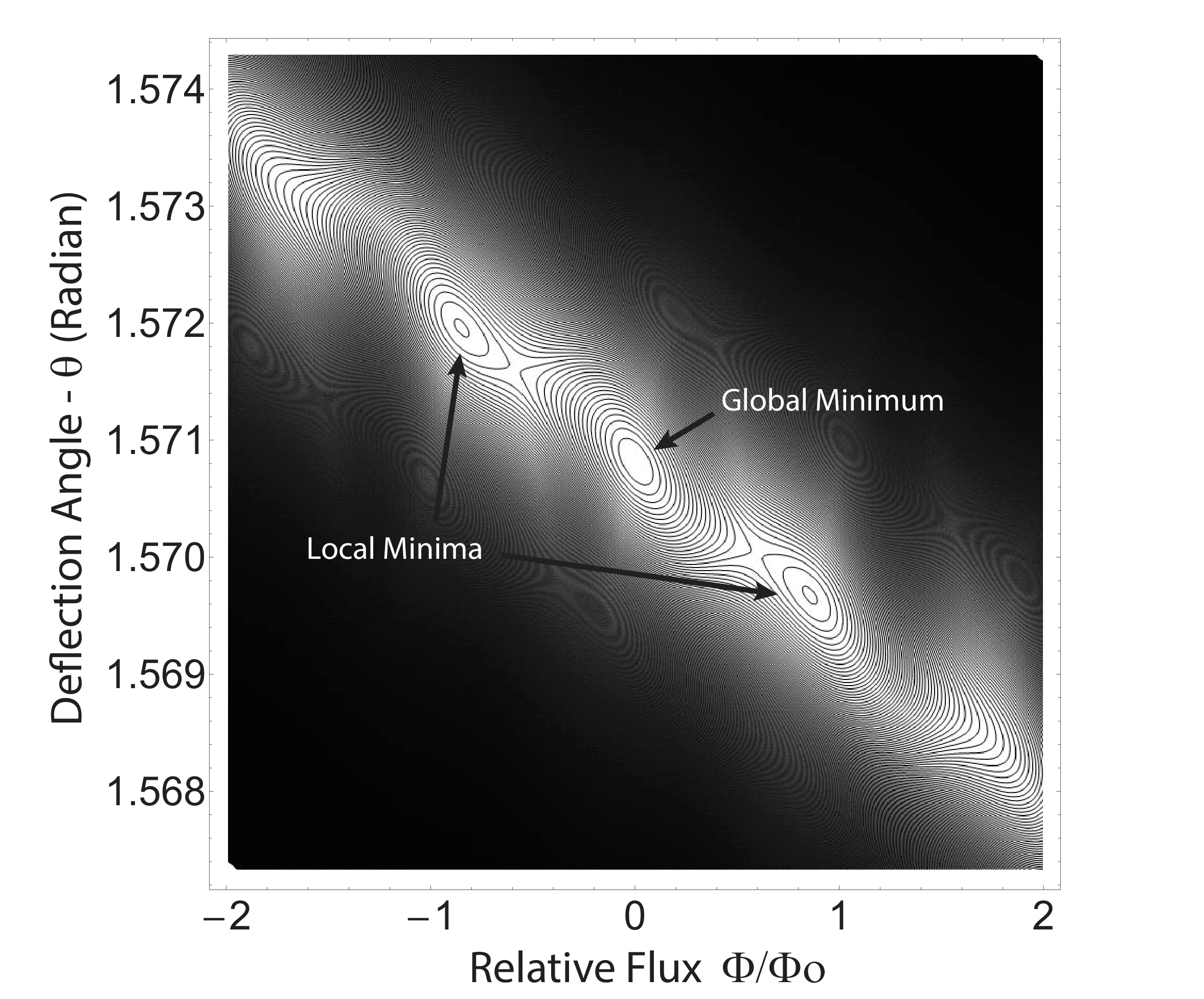}
\caption{\label{fig2} \emph{A contour plot indicating location of a two-dimensional global potential energy minimum at ($n\Phi_{0}=0,\theta^{+}_{n}=\pi/2$) and the local minima when the cantilever equilibrium angle $\theta_{0}=\pi/2$, $\omega_{i}=2\pi\times 12000$~rad/s, $B_{x}=5.0\times 10^{-2}$T. The contour interval in units of frequency ($E/h$) is $\sim3.9\times 10^{11}$Hz.}}
\end{center}
\end{figure}
\end{center}

Consider a flux-qubit-cantilever made of niobium which is a type-II superconductor with a transition temperature of about $9.26~K$. Further, consider a square cross-section of this superconducting material with edge thickness $t= 0.5 \mu$m,  $l=6\mu$m, $w=4\mu$m ($A=l\times w$). For these dimensions, the mass of the cantilever is $3.64\times10^{-14}$Kg and the moment of inertia is $I_{m}\simeq7.28\times10^{-25}$ Kg~m$^{2}$. The critical current of Josephson junction $I_{c}=5\mu$A, capacitance $C= 0.1$pF and self-inductance $L=100$pH which are of the same order as described in reference~\cite{fried}. The quantity $\beta_{L}=2\pi L I_{c}/\Phi_{o}\simeq 1.52$. Consider the intrinsic frequency of the cantilever $\omega_{i}=2 \pi \times12000$~rad/s. For an equilibrium angle $\theta_{0}=\theta^{+}_{n}=\cos^{-1}[n\Phi_{o}/B_{x}A]$ there exists a single global potential energy minimum. If we consider $n=0$ and $B_{x}=5\times 10^{-2}$T the global potential energy minimum is located at ($n\Phi_{0}=0,\theta^{+}_{n}=\pi/2$). For parameters described above  $\omega_{\phi}\simeq2\pi\times7.99\times10^{10}$~rad/s, $\omega_{\delta}=2\pi\times25398.1$~rad/s and $\kappa=0.012$~A. The eigen frequencies of the flux-qubit-cantilever are $\omega_{X}\simeq2\pi\times7.99\times10^{10}$~rad/s and $\omega_{Y}=2\pi\times21122.5$~rad/s. A contour plot of the two-dimensional potential energy of the flux-qubit-cantilever indicating a two-dimensional global minimum located at ($n\Phi_{0}=0,\theta^{+}_{n}=\pi/2$) and two local minima is shown in Fig.~\ref{fig2}.
 Even if we consider intrinsic frequencies to be zero, the restoring force is still nonzero due to a finite coupling constant. For $\omega_{i}=0$, the angular frequencies are $\omega_{\phi}\simeq2\pi\times7.99\times10^{10}$~rad/s, $\omega_{\delta}=2\pi\times22384.5$~rad/s, $\omega_{X}\simeq2\pi\times7.99\times10^{10}$~rad/s and $\omega_{Y}=2\pi\times17382.8$~rad/s. These frequencies can be tuned by varying the external magnetic field and by changing the dimensions of the cantilever, which alters its mass, moment of inertia and self-inductance.
\begin{center}
\begin{figure}
\begin{center}
\includegraphics[scale=0.24]{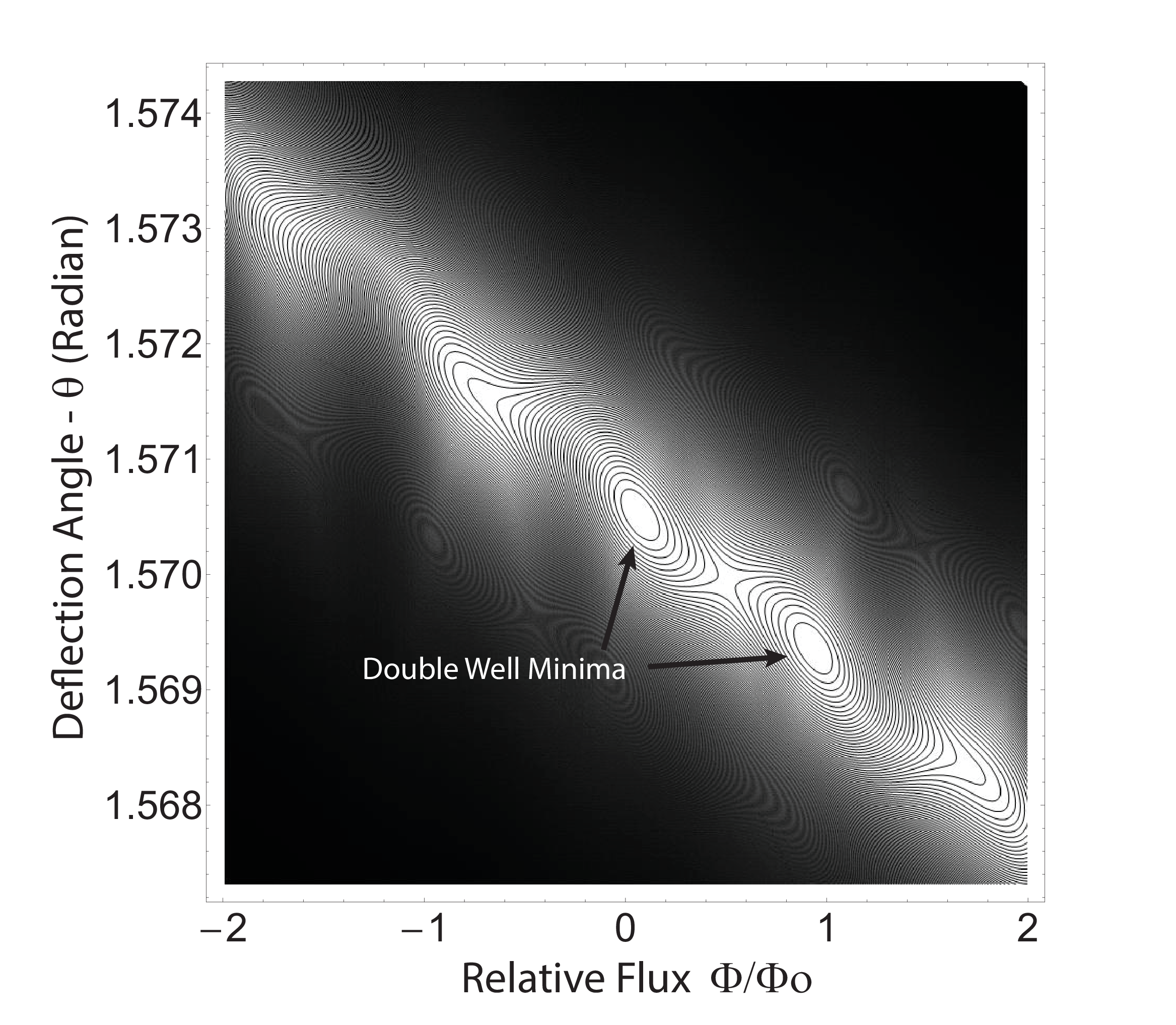}
\caption{\label{fig3} \emph{A contour plot indicating location of two-dimensional potential energy minima forming a symmetric double well potential when the cantilever equilibrium angle $\theta_{0}=\cos^{-1}[\Phi_{o}/2B_{x}A]$, $\omega_{i}=2\pi\times 12000~rad/s$, $B_{x}=5\times 10^{-2}$T. The contour interval in units of frequency ($E/h$) is $\sim4\times 10^{11}$Hz.}}
\end{center}
\end{figure}
\end{center}

The potential energy of the flux-qubit-cantilever near its equilibrium angle corresponds to a symmetric two-dimensional double well (\emph{i.e.} two global minima for $m\Phi_{o}<B_{x}A<(m+1)\Phi_{o}$) if the equilibrium angle of the cantilever is chosen such that $\theta_{0}=\cos^{-1}[(2n+1)\Phi_{o}/2B_{x}A]$, provided $\omega_{i}$ is less than or of the order of the first term of $\omega_{\delta}$ (Eq.~\ref{eq:omega}) \emph{i.e.} $(B^{2}_{x}A^{2}-n^{2}\Phi^{2}_{o}/I_{m}L)^{1/2}$. For a two-dimensional double well potential, the left potential well is located near ($n\Phi_{o},\theta^{+}_{n}$) and the right potential well is located near ($(n+1)\Phi_{o},\theta^{+}_{n+1}$). Any variation of the equilibrium angle around $\cos^{-1}[(2n+1)\Phi_{o}/2B_{x}A]$ introduces an asymmetry in the two-dimensional double well potential. The asymmetry in the two-dimensional double well potential can also be produced by tilting the magnetic field direction. Consider $n=0$ such that $\theta_{0}=\cos^{-1}[\Phi_{o}/2B_{x}A]$; therefore, at equilibrium the flux-qubit-cantilever is biased at half of the flux quantum ($\Phi_{o}/2$). A contour plot indicating a two-dimensional symmetric double well potential is shown in Fig.~\ref{fig3}. Consider for the double well potential, the non-separable ground states of the left and the right potential well to be $|\alpha\rangle_{L}$ and $|\alpha\rangle_{R}$, respectively. The barrier height between the two wells of the two-dimensional double well potential which is less than $2E_{j}$, reduces when $\omega_{i}$ is increased. The barrier height controls the tunneling between the potential wells and it can also be tuned through an external magnetic flux applied to the dc-SQUID of the flux-qubit-cantilever. When the tunneling between the wells is introduced, the ground state of the flux-qubit-cantilever is $|\Psi\rangle_{E}=[|\alpha\rangle_{L}+|\alpha\rangle_{R}]/\sqrt{2}$. The state $|\Psi\rangle_{E}$ is an entangled state of the magnetic flux and the cantilever deflection. The state  $|\Psi\rangle_{E}$ can be realized by cooling the flux-qubit-cantilever to its ground state.

A special case of interest of the flux-qubit-cantilever when the intrinsic frequency $\omega_{i}$ of the flux-qubit-cantilever is zero, the potential energy $V(\Phi,\theta)=(\Phi-B_{x}A\cos(\theta))^{2}/2L + E_{j}\left( 1-\cos(2\pi\Phi/\Phi_{o})\right)$, has multiple two-dimensional global minima forming a lattice. For $m\Phi_{o} < B_{x}A < (m+1)\Phi_{o}$, integer $m \geq 0$, the minima of potential energy are located at ($n\Phi_{0},\theta^{+}_{n}$) and ($n\Phi_{0},\theta^{-}_{n}$), where $n=-m,-m+1,..0..,m-1,m$.

\section{Superconducting-Loop-Oscillator}
\begin{center}
\begin{figure}
\begin{center}
\includegraphics[scale=1.35]{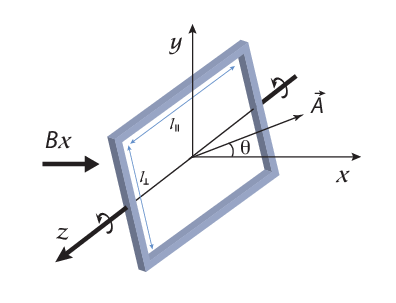}
\caption{\label{loop} \emph{A superconducting-loop-oscillator with its axis of rotation along the $z$-axis consists of a closed superconducting loop without a Josephson Junction. The superconducting loop can be of any arbitrary shape.}}
\end{center}
\end{figure}
\end{center}

Here a superconducting-loop-oscillator which is a variant of the flux-qubit-cantilever is considered, which consists of a superconducting loop without a Josephson junction, as shown in Fig.~\ref{loop}. The axis of rotation of the superconducting loop coincides with the $z$-axis. Both ends of the loop axis are mounted on a substrate (which is not shown in Fig.~\ref{loop}). The closed superconducting loop can be of any arbitrary shape. The Hamiltonian of such an oscillator is a special case of the Hamiltonian of Eq.~\ref{eq:hamiltonian} if the Josephson potential energy term is zero. Therefore, the Hamiltonian of the superconducting-loop-oscillator can be written as
\begin{eqnarray}\nonumber
\label{eq:hamiltonianloop}
H&=&\frac{p^{2}_{\theta}}{2 I_{m}}+\frac{1}{2}I_{m}\omega^{2}_{i}(\theta-\theta_{0})^{2}\\&& +\frac{(n\Phi_{o}-B_{x}A\cos(\theta))^{2}}{2 L}
\end{eqnarray}
it is assumed here that a definite value of the quantized flux $n \Phi_{o}$ is threading through the loop, and the oscillation of the loop does not excite or de-excite the magnetic flux quantum state. Such an approximation is valid if the frequency of oscillation of the superconducting loop is much less than the frequency equivalent of a energy difference between different flux quantum states. Since the magnetic flux trapped in a superconducting loop remains constant, therefore the potential energy profile of the superconducting loop oscillator is a one-dimensional function (for a fixed value of $n$) $V(\theta)=\frac{1}{2}I_{m}\omega^{2}_{i}(\theta-\theta_{0})^{2}+\frac{1}{2 L}(n\Phi_{o}-B_{x}A\cos(\theta))^{2}
$. In the presence of an external magnetic field $B_{x}$, the superconducting loop acts as an oscillator even if its intrinsic frequency is zero. Here the area $A=l_{\|}\times l_{\perp}$, where $l_{\|}$ and $l_{\perp}$ are the dimensions of the loop as shown in Fig.~\ref{loop}.

Consider $B_{x}A < n\Phi_{o}$ and $\theta_{0}=0$. In this case the potential energy minimum is located at $\theta=0$, and the restoring couple acting on the superconducting-loop-oscillator for a small amplitude oscillation near the minimum is $\tau_{\theta}=-\left(I_{m}\omega^{2}_{i}+(n\Phi_{o}-B_{x}A)B_{x}A/L\right)\theta$. The restoring torque which is proportional to the angular displacement reflects, the harmonic nature of the potential near the minimum. A small amplitude oscillation frequency in such a harmonic potential is calculated as
\begin{equation}
\label{eq:onedh}
\omega_{n(H)}=\left(\omega^{2}_{i}+\frac{(n\Phi_{0}-B_{x}A)B_{x}A}{I_{m}L}\right)^{1/2}
\end{equation}
where the subscript $n(H)$ of $\omega_{n(H)}$ signifies the harmonic potential corresponding to the $nth$ quantum of the trapped magnetic flux. The $\omega_{n(H)}$ is nonzero even if intrinsic frequency $\omega_{i}$ is zero. On the other hand, for the case if $B_{x}A > |n\Phi_{0}|$, $\omega_{i}=0$ and $\theta_{0}=0$, there exist two local minima of $V(\theta)$ located at $\theta^{+}_{n}$ and $\theta^{-}_{n}$. In this case $V(\theta)$ corresponds to a one-dimensional double well potential. A deviation of $\theta_{0}$ around zero produces an asymmetry in the double well potential. The asymmetry in the double well potential can also be produced by tilting the magnetic field direction. A small amplitude oscillation frequency around each minimum of the double well is calculated as
\begin{equation}
\label{eq:oneddw}
\omega_{n(DW)}=\left(\frac{B^{2}_{x}A^{2}-n^{2}\Phi^{2}_{0}}{I_{m}L}\right)^{1/2}
\end{equation}
where a subscript $n(DW)$ signifies a double well potential corresponding to the $nth$ quantum of the trapped magnetic flux. The barrier height which is $V_{B}=(n\Phi_{0}-B_{x}A)^{2}/2L$ for $\omega_{i}=0$ decreases when $\omega_{i}$, is increased.
\begin{center}
\begin{figure}
\begin{center}
\includegraphics[scale=0.4]{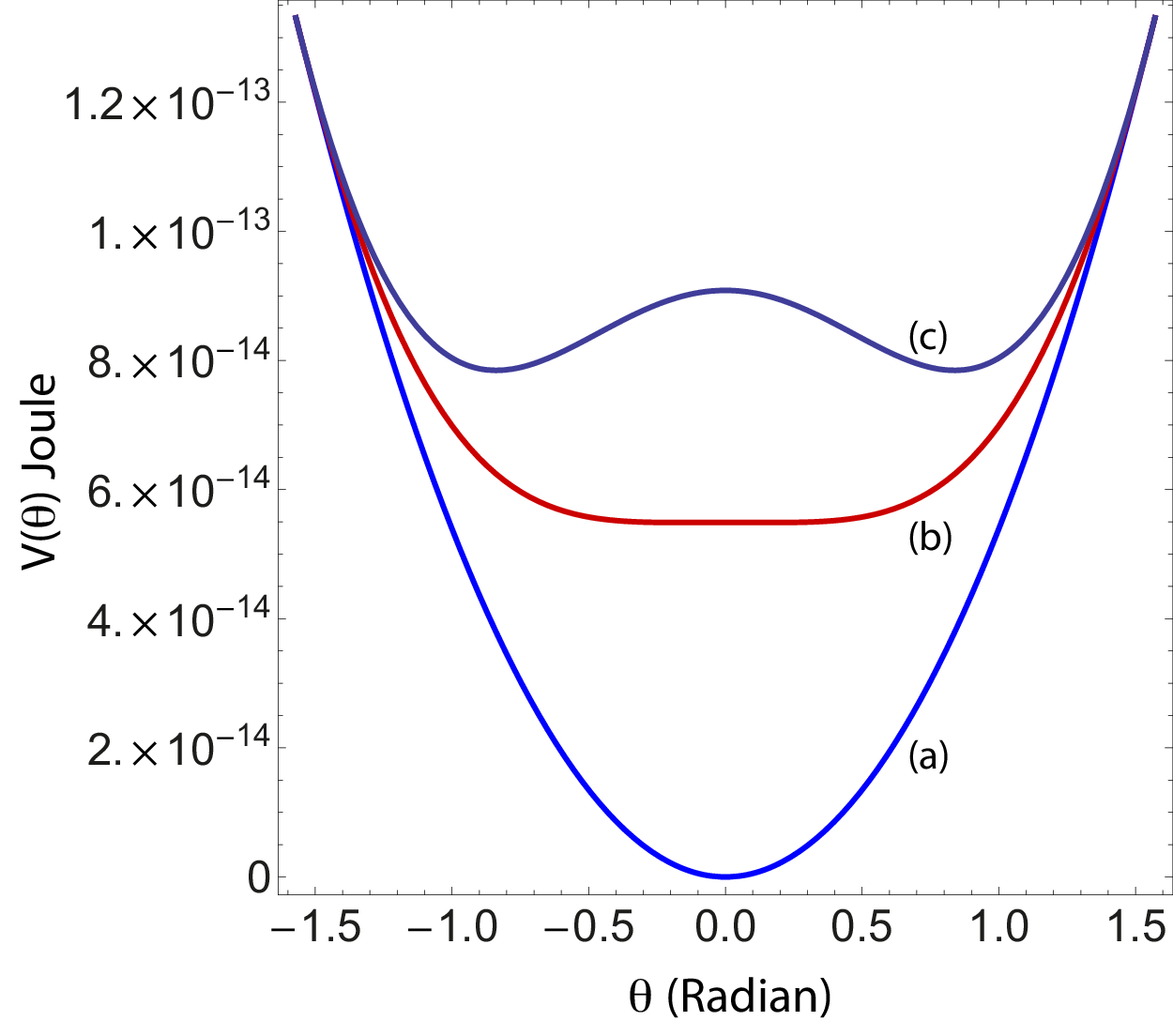}
\caption{\label{loopplot} \emph{The potential energy profile of the superconducting-loop-oscillator when the intrinsic frequency is 10~kHz. (a) For external magnetic field $B_{x}=0$, a single well harmonic potential near the minimum is formed. (b) $B_{x}=0.035~$T (c) for $B_{x}=0.045~$T, a double well potential is formed.}}
\end{center}
\end{figure}
\end{center}

The one-dimensional potential $V(\theta)$ can be gradually modified from a single harmonic well potential to a double well potential by increasing the external magnetic field $B_{x}$. Such a gradual modification of the potential energy for a single flux quantum (n=1) and $\omega_{i}=10~$kHz is shown in Fig.~\ref{loopplot}, where in Fig.~\ref{loopplot} (a) a single harmonic potential is formed for an external magnetic field $B_{x}=0~$T, Fig.~\ref{loopplot} (b) $B_{x}=0.035~$T and Fig.~\ref{loopplot} (c) a single harmonic potential is modified into a double well potential for $B_{x}=0.045~$T. For these plots a niobium superconducting-loop-oscillator of dimensions $l_{\|}=6\mu$m, $l_{\perp}=5\mu$m is considered, where niobium has a square cross-section of edge thickness $0.5 \mu$m.
The total mass, the moment of inertia along the axis of rotation and the self-inductance of the superconducting-loop-oscillator are $5.14 \times 10^{-14}$~kg, $2.73 \times 10^{-25}$Kgm$^{2}$ and $10$pH, respectively.
A gradual modification from a single well potential to a double well potential has an important application in realizing a macroscopic quantum superposition state of the superconducting-loop-oscillator. Consider that the superconducting-loop-oscillator is prepared in the ground state of a single well harmonic potential when the external magnetic field is zero. When the external magnetic field is increased adiabatically, the superconducting-loop-oscillator remains in the ground state of its instantaneous potential. Such an adiabatic increase of the magnetic field, gradually keeps the superconducting-loop-oscillator in the ground state of the double well potential. The ground state of the double well potential corresponds to a macroscopic superposition of distinct configurations. However, the potential near its minimum relaxes for an intermediate value of an external magnetic field as shown in Fig.~\ref{loopplot} (b), and is the regime where the external magnetic field has to be increased very slowly.

\section{Concluding Remarks}
In this paper a macroscopic quantum oscillator (flux-qubit-cantilever) is proposed, which exhibits a natural coupling between the magnetic flux and the cantilever. The coupling constant can be varied by the external magnetic field. The potential energy is adjusted through the cantilever equilibrium position and the external magnetic field. The ground state of the flux-qubit-cantilever corresponds to an entangled quantum state of the magnetic flux and the cantilever deflection. Furthermore, a variant of the flux-qubit-cantilever is also proposed, which is a superconducting loop without any Josephson junction. This superconducting-loop-oscillator provides a way of generating a macroscopic quantum superposition of distinct configurations. Both the proposed systems can be experimentally implemented and provide a novel way of generating quantum entangled states and quantum superpositions of macroscopic observables. In both the cases, the novel quantum phenomenon occurs in the ground state, which provides an experimental advantage.

\begin{acknowledgments}
Author is thankful to Prof. Arvind for useful comments on this paper.
\end{acknowledgments}

\bibliography{fqc1}

\end{document}